# Reoccurring patterns in hierarchical protein materials and music: The power of analogies

*Tristan Giesa[1,2], David I. Spivak[3] and Markus J. Buehler[1,4*]*

[1] *Laboratory for Atomistic and Molecular Mechanics, Department of Civil and Environmental Engineering, Massachusetts Institute of Technology, 77 Mass. Ave. Room 1-235A&B, Cambridge, MA, 02139, USA*
[2] *Department of Mechanical Engineering, RWTH Aachen University, Templergraben 55, 52056 Aachen, Germany*
[3] *Department of Mathematics, Massachusetts Institute of Technology, 77 Mass. Ave, Cambridge, MA, 02139, USA*
[4] *Center for Computational Engineering, Massachusetts Institute of Technology, 77 Mass. Ave. Room 1-235A&B, Cambridge, MA, 02139, USA*

**\* Corresponding author, E-mail:** mbuehler@MIT.EDU

**ABSTRACT:** Complex hierarchical structures composed of simple nanoscale building blocks form the basis of most biological materials. Here we demonstrate how analogies between seemingly different fields enable the understanding of general principles by which functional properties in hierarchical systems emerge, similar to an analogy learning process. Specifically, natural hierarchical materials like spider silk exhibit properties comparable to classical music in terms of their hierarchical structure and function. As a comparative tool here we apply hierarchical ontology logs (olog) that follow a rigorous mathematical formulation based on category theory to provide an insightful system representation by expressing knowledge in a conceptual map. We explain the process of analogy creation, draw connections at several levels of hierarchy and identify similar patterns that govern the structure of the hierarchical systems silk and music and discuss the impact of the derived analogy for nanotechnology.

## 1. Introduction

Comprehension by analogies is a widely applied concept in science and education [1,2,3,4]. Successful educational strategies comprise features such as 'constructivist learning environments' that challenge the view that scientific and mathematical knowledge is static, independent from our minds, and represents a universal truth [5]. In fact, this knowledge serves as a mediator resulting from human inquiry. Children in school become introduced to sciences such as mathematics, physics, chemistry and biology *via* the link to structures and concepts they are more likely to be familiar with. For instance, atom and electron interactions are represented by a model that resembles galaxy structures - the Bohr model - or animal cells are represented by factories, see *e.g.* [6,7]. While most people may agree that it does not represent reality even closely, the analogy provides a sufficiently thorough understanding of general mechanisms that take place on the scales of Angstroms and nanometers.

By systematic abstraction and the deduction of analogy steps, the process of building an analogy itself helps to boost discretion about the important properties and parameters of the inquired system - or at least to ask essential questions [8,9]. A rigorous methodology to formulate and categorize these analogies can be provided by so-called ontology logs (olog), based on category theory [10]. Category theory originates from the 1940s as a mathematical concept in topology [11] and has recently been used in broader contexts to identify patterns in language, physics, philosophy, and other fields in a mathematical framework [12,13,14,15]. Categorical algebras consist of objects and arrows which are closed under composition and satisfy conditions typical of the composition of functions [16]. In a linguistic version, category theory and ologs in particular describe the essential features of a given subject and represent a powerful method to store and share data, knowledge, and insights in structure and functionality.

Equivalent to the learning and understanding process applied in school physics and mathematics, the methodology and hence the above described advantages of analogical thinking can be adapted to the relatively novel field of materiomics [17]. Recent studies linked for example mechanical properties of protein networks to communication networks [18], active centers in proteins to top predators and top managers [19], and death of living organisms to the strength of solids [20]. Furthermore, the connection between grammar and protein structure has been elaborately studied [21], even in the context of category theory [22].

Here, through the example of spider silk and classical music, the construction of simple analogies and their accomplishment to collocate a broad picture of materials structure and function is showcased. Thereby, similar patterns in silk and music which trace back to their hierarchical build-up are identified. The aim of this study is the institution of category theoretic tools as a rigorous and comprehensive means to systematically depict and communicate hierarchical structure-function relationships for applications across fields in engineering, science and the arts.

## 2. Methods

### 2.1 Ologs

Category theoretic analysis and transformations of syntactic structures have been introduced by Chomsky in 1957 [23]. For formal language theory a well-known transformation is for example the left part transformation from non-left-recursive context-free grammars to context-free Greibach normal form where the syntactic structure is preserved during the transformation [24]. These structure preserving transformations are *morphisms* (or *functors*) between objects and arrows among categories and constitute the essential operator to form analogies. The linguistic categorical objects in ologs are sets, and the arrows represent unique functions between the objects. Ologs are embedded into a database framework and thus are easily implemented in object based computer languages.

The actual benefit of an olog is - due to a rigorous mathematical background - its unambiguous way to store and share data, knowledge, and insights in structure and functionality within a single research group and also among many disparate research groups and different fields in science and engineering. Ologs offer means to reveal the origin of the described system property and to connect them to previous results or other topics and fields. For biological materials it is crucial to elucidate the principles from which the superior macroscopic functionality arises in order to define the hierarchical structure-function relationships or even synthesize them. These insights can be gained on the one hand from the category theoretic analysis of protein materials by describing the emergence of functionality from first principles, *e.g.* on the basis of fundamental interactions between building blocks. On the other hand, the use of systematic analogies with the help of *functorial* relations supports the researcher in formulating these structure-function relationships in an abstract way, ensuring the ability to connect disparate ologs. Furthermore, by the use of ologs for knowledge creation by shared conceptual models an educational application is feasible [25].

**2.2 Hierarchical ologs**

Hierarchical ologs yield similar mathematical features as conventional ologs but are designed to improve the ability to overview the build-up of a hierarchical system by compiling the subunit sets together with sets of superior structural units. As an example of the typical features found in hierarchical ologs we analyze a distinct feature of linguistics, specifically the structure of a sentence which is formed of words, Figure 1a. Words consist of phonemes, the smallest pronounceable segments comprising one or more letters [26,27]. Hence, these phonemes form the 'building blocks' for spoken language and are categorized by distinctive articulatory features, *i.e.* the description of how the sound is mechanically formed in the body [28]. Whether or not a feature is active for a certain phoneme is indicated by a binary number (alternatively by a + or – sign). The voice laryngeal feature, for instance, determines whether a sound is formed including the vocal folds (1) or not (0) [29,30]. The 'r' in 'theory' contains as voice laryngeal feature a (1), while the 'e' does not (0).

Each path is constrained to represent a unique function between the instances of the sets. Consequently, each building block can only be uniquely assigned to its higher units by constructing pairs of building blocks and their higher structural units respectively, see Figure 1b. Set **A** associates units from set **B** and set **C** together with their positions, an elegant way to maintain functional relationships within the category. The green checkmarks indicate commutative paths (a kind of *isomorphism)* in the olog where, starting from the same instance of a set, two distinct paths point to the same instance of another set. For further details on *isomorphisms* and other features such as *limits* and *colimits* in ologs see [10,18,31]. The way the olog is represented in Figure 1b correlates one-to-one to a computer implementation.

Beneficial to the possibility to overview and clearly identify the underlying hierarchical structure of the system, we introduce hierarchical ologs, Figure 1c. They yield absolutely identical information but a well-arranged notation. In this context, an increased emphasis on the discovery of the structural makeup of systems and how it relates to the emergence of specific properties can provide an important educational feature. Set **A** of Figure 1b is now inherently included and the dashed box represents a state of the system, *i.e.* all words in the former set **B** with their current phonetics and binary information (comparable to a 'snapshot'). Often, for instance in the case of phonemes in words, a higher structural unit (the word) is formed by a sequence of subunits (phonemes). This information is conveyed by a simple dashed 'hierarchy-box': the word box surrounds the phoneme box, indicating a hierarchical construction. In other cases, the structure may be arbitrary, but always describable by graph-theoretic tools. This holds true in the case of proteins, which are arrangements of (*i.e.* hierarchically constructed from) amino acides  Paths combining 'inner' and 'outer' information, *e.g.* the new 'double arrows' and former commutative paths, are automatically commuting and no additional checkmarks are needed.

Hierarchical ologs can be described easily using category theory. As explained in [10], an olog is a category (or more precisely a sketch). A category theory expert can understand our definition of a *hierarchical olog* as a category $C$ equipped with a subcategory $H$ with the same set of objects, Ob($C$)=Ob($H$), and such that $H$ has the structure of a tree. A morphism $F: (C,H) \to (C',H')$ of hierarchical categories, which we here abuse notation in calling a *functor*, consists of a functor $F: C \rightarrow C'$ such that $F(H) \subseteq H'$ and $F$ preserves the root. A state of a hierarchical olog is just a set-valued functor on $C$; we can denote sequences, graphs, etc. by a simple modification of the olog, which again we abuse notation in eliding.

**3. Results and discussion**

As an application for the generation of an analogy, we construct an olog that reflects the hierarchical structure found in protein materials such as spider silk, Figure 2. In order to form the analogy to music we must determine a way to dissect the structure to its basic constituents. A generally advisable approach is the definition of building blocks of the systems first. Depending on the level of abstraction, these building blocks can be of real nature, *e.g.* phonemes, or of abstract nature, *e.g.* the 'lifeline' in beta-helices [18]. Here, proteins assemble out of their building blocks, amino acids, whereas we define the building blocks of music as sound waves (sine, triangular, sawtooth, *etc.*, [32,33]) that are assumed to assemble via stacking, *i.e.* without any additional information about amplitude, frequency or pitch.

In a second step, we define the superior structural units and indicate how they are related to their basic constituents. Bonds affiliate amino acids into groups and thereby a polypeptide is a linear chain of two or more amino acids connected by a (peptide) bond. Each bond within the polypeptide has as starting point and end point an amino acid and hence they represent a subset of amino acid groups which are, in contrast to polypeptides, not necessary a linear chain of amino acids but can assemble in more complex structures (describable *via* a 3d graph). In an analogous way we define the creation of musical structural assemblies where stacked groups of sound waves are called a tone.

So far, the relations only concern structure terminology and the question how functionality can be addressed remains open. Proteins, *i.e.* groups of one or more polypeptides, fold into secondary structures which are crucial to their properties and functionality, see *e.g.* [34]. Hence, a precursor to the assignment of sequence-structure-function relationships is the sequence-structure identification by experiment and computational studies. Such knowledge-based assignments have already been part of preceding inquiries [35,36].The information gathered from these studies, for example the sequence and environmental conditions that lead to distinct structural assemblies, then become data in the olog. Nanocomposites consist of proteins positioned in certain secondary structures (*e.g.* alpha helix, beta sheet or amorphous phase in spider silk [37,38,39]; in any case defined by a graph structure) of a specific size that assemble into higher level networks. The shear strength of these secondary structures, information stored in 'a shear strength', is directly related to properties such as size and arrangement [40,41,42]. Similarly, the variation of frequency and amplitude of the stacked waves leads to the formation of the functional unit 'a note' defined by its property 'a pitch' [43]. The pitch corresponds to the audibility which then determines together with duration, loudness and timbre the functional properties of a chord (grouping of intervals into categories such as thirds, fourths, etc.) which assemble into harmonically stable riff structures [44,45,46,47,48,49,50]. Here we identify a major potential of hierarchical ologs. As all chords in the riff are assembled in a weighted graph structure, the information that riffs are made of rhythmic arrangements of chords is inherently included. This designates a novel way of writing music sheets; where chords represent graph nodes connected to their nearest neighbors by edges where the edge length (or weight) directly correlates to the length of the chord.

Apart from the hierarchical levels, a subset of proteins, the enzymes, are included to exemplify the procedure to include 'hierarchical interaction'. Subsets of higher hierarchical levels contain distinguished members whose functionality is based on lower level architecture. Thus, a function relates a certain property of a higher hierarchical structure, here active centers of enzymes, to an element of a lower level structure, a group of amino acids. Similar to the enzyme-protein relation subsets of chords also include group members with distinguished functional meaning [47,51]. Here, the major chords, as interval special functional class of chords, has as distinguished member the root, the base on which a triadic chord is built. This kind of relation is typical for all kinds of hierarchical organizations, *e.g.* in primate groups [52]. After assembling and relating these insights, a challenge that can be overcome by multiscale studies including graph theoretic tools, a deduction to fields which show an equivalent hierarchical buildup by *functors* is possible, Figure 2. Thereby, the relations and thus the functionality within the category are maintained and the two seemingly disparate fields display their intrinsic connection. In this example the *functorial* transformation is an *isomorphism* meaning that the positions of boxes and arrows are the same in both systems; thus it requires no further clarification.

Apart from the simple description of structural details (graph theory would actually provide sufficient means for that), ologs also reveal system properties in a category-theoretic framework. Such a property is for example the H-bond clustering found in protein structures like spider silk [41]. Geometric confinement of protein materials at the nanoscale leads to the rupture of clusters of 3-4 hydrogen bonds in the β-sheet structures and thus to an optimized shear strength. This is shown in Figure 3 where a functional property of the cluster, the shear strength, is related to a structural condition, the geometric confinement. This olog is based on the insights gained from multiscale computational studies [41]. Similar cluster strategies can be found in music. Chords consisting of 'consonant' frequencies are considered to be innately pleasant to humans and even some animals, *i.e.* these frequencies belong to an (equally tempered) linear pitch space where pairs of frequencies $f_i, f_j$ follow the approximate relation $c|\log_2(f_i/f_j)| \epsilon\ N$ with $c = 12$ [53,54]. We check this condition for consonance by determining the matrix $[x_{ij}]$ and checking whether the entries have indeed integer values. This physically quantifiable ratio results in the *empirically* quantifiable sensation revealing emotions, data that can be incorporated in our olog [55,56]. The major difference between materials sciences and social or artistic sciences is disclosed here: material properties are usually unambiguous (well-defined) and objective whereas the evaluation of artistic properties is subjective. The present *functorial isomorphism* allows the correlation of these data and subsequent statistical analyses may reveal additional insights that then lead back to the original system – an advantageous approach to recycle knowledge of well-studied systems such as music for novel applications.

Concerns could relate to the fact that the secondary structure of proteins is often not deterministic, *i.e.* the same protein can fold into more than one structure which endangers the unambiguity of the functional relationships. The same holds true for music, where for example pitch and timbre are sometimes ambiguous [57]. This has to be addressed by determining the environmental conditions that uniquely specify the protein's secondary structure, thus defining a protein grammar [21] or similar, again by experiment and computational modeling.

Both ologs, Figure 2 and 3, are part of a bigger olog which would describe the material system spider silk (or classical music respectively) in total. Addressing the challenge to complete the olog immediately, it is straightforward to start with subunits as presented here. The assembly of a bigger system would work for example with the set 'a bond'/'a stack', which is shared in both ologs and hence serves as an attachment point to combine them. Table 1 summarizes key structures and functions where connatural hierarchical ologs could be designed to uncover more analogies and then be attached to the existing parts. For instance, higher order structural assemblies such as nanocomposites/riffs gain functional importance by pattern building. In proteins this may relate to the repetition of secondary structural units and their overall confinement which ensures macroscale functionality, for example semi-amorphous phase and β-sheet domains in spider silk that provide a superior toughness by confinement to a fibril size of around 50 nm [58]. The corresponding pattern in music is the formation of chord sequences into riffs and phrases *via* syntactic structures that provide musical tension, an important functional focus in music [59,60]. Yet another functional commonality of the two systems silk and music is related to the damage tolerance behavior. Localized defects in spider webs do not effect overall mechanical functionality [61] while the deletion of certain chords in a chord sequences do not affect the tonal coherence and hence the functionality [62]. Note that in hierarchical systems functionality is generally obtained by structural arrangement, *e.g.* clustering or stacking, and hence it is mostly impossible to separate structure and function. Therefore, the structure and function dimension in Table 1 overlap.

### 4. Conclusion

Here we introduced hierarchical ologs to describe typical natural hierarchical systems such as language, biological materials and music and draw intrinsic connections between the underlying structures. We showed that in protein materials such as spider silk hierarchical structures identical to classical music can be identified and properly documented by means of hierarchical ologs. Similar to an analogy learning process this method may on the one hand serve as a guide to construct ologs for data and knowledge sharing in research groups and on the other hand as a method to utilize analogies to teach structure and functionality of hierarchical material systems. Conceptually organized like Wikipedia (http://www.wikipedia.org/), but with a substantiated mathematical background, ologs may provide a powerful academic and scientific tool to categorize, organize, relate and share insights gained during research. Specifically, analogy building as a momentous instrument for human understanding and education may be formalized by the use of category theory based ontologies. The ultimate strength of this tool relates to the enforcement of rigor during the analogy building process. Each term and concept defined in one system needs to be precisely related to their analogy counterpart to obtain a structure preserving transformation *via functors*. This characteristic feature of ologs sets the fundament for their superiority over common ontological or heuristic approaches.

**Acknowledgements**: We acknowledge support from AFOSR and DOD-PECASE (funded by ONR grant # N00014-10-1-0562). Additional support was received from the German National Academic Foundation (Studienstiftung des deutsches Volkes) and ONR grant # N00014-10-1-0841.

**Table and Table Caption**

**Table 1 | Overview over similarities between spider silk and music.** Beyond the patterns shown in Figure 2 and Figure 3 other structural and functional similarities between spider silk and music can be identified.

| | General Property | Silk | Music |
|---|---|---|---|
| **Structure** | Assembly of building blocks | Amino acids assemble into polypeptides via polypeptide bonds | Sound waves are stacked and interfere |
| | Assembly of single units | Polypeptides assemble via covalent and weak bonds and form secondary structures | Sound waves with different frequency, amplitude and pitch form notes (instrument) |
| | Assembly of functional units | Silk protein is formed in a stable structure dependent on solvent condition and ionization state | Sound wave of consonant frequency form chords on the equally tempered scale |
| | Assembly of functional structure | Alanine rich repeat units form beta-sheets with high strength whereas glycine rich repeat units form extensible semi-amorphous phases; Repetition of functional units creates nanocomposites | Harmonic sequences consist of the three main functions (tonic, sub-dominant and dominant); Sequence/Repetition of chords creates a melody riff |
| **Function** | Upscaling of functionality | Nanoconfinement of composite structure ensures functionality (high strength, extensibility and toughness) on the macroscale | Phrases and climaxes within the music ensure musical tension, functional dependency of chord sequence |
| | Damage Tolerance | Localization of deformation upon loading provides spider webs with robustness, damage mitigation, and superior resistance by nonlinear material behaviour | According to the dependency structure single chords may be removed from or entered into the sequence without affecting the harmonic function |

**Figures and Figure Captions**

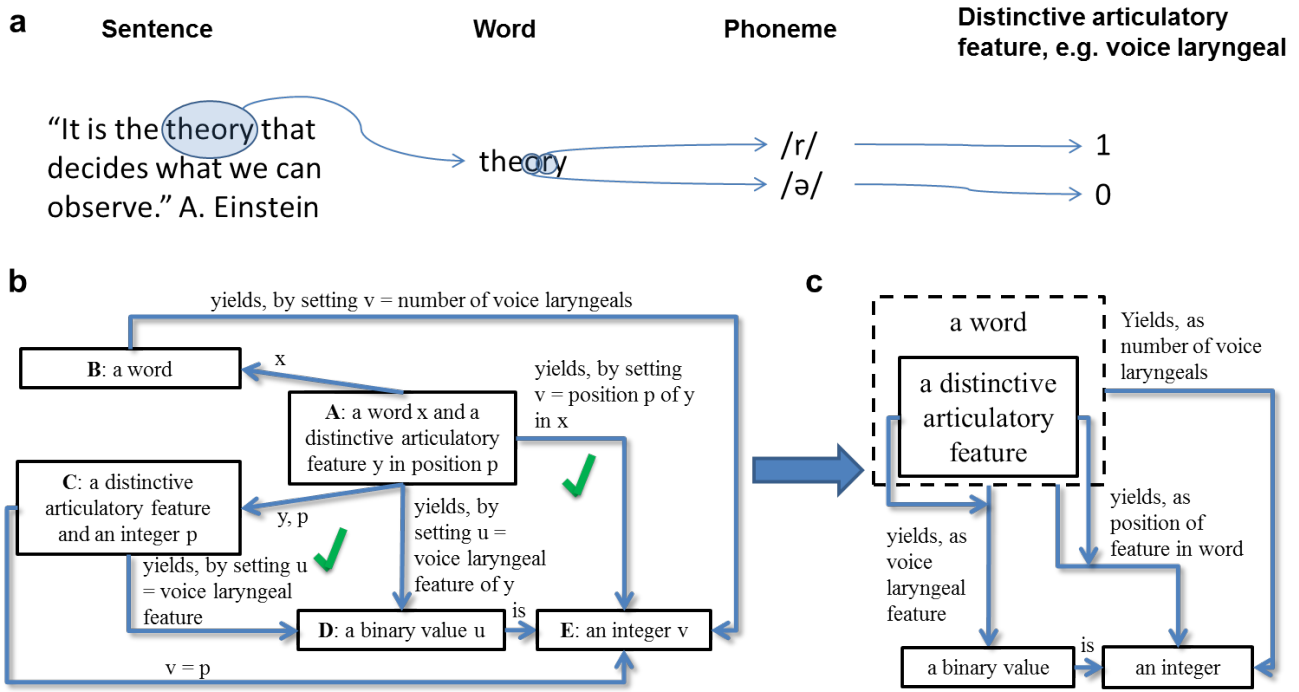

**Figure 1 | Linguistic structure and category theory. a**, The hierarchical buildup of linguistics can be seen when analyzing the structure of a sentence. It is formed by words which itself are formed by phonemes, the smallest pronounceable segments of a word. These phonemes are categorized by distinctive articulatory features, *i.e.* the description of how the sound is mechanically formed in the body. Whether or not a feature is active for a certain phoneme is indicated by a binary number. The voice laryngeal feature determines whether a sound is formed including the vocal folds or not. Note, that the paths A➔D and A➔C➔D as well as A➔E and A➔C➔E commute as indicated by the green checkmarks. No other paths commute. **b**, Olog describing the situation where every word consists of phonemes, hence they form the 'building blocks' for spoken language. An appropriate way to assign the building blocks to its higher units is a representation of sets of words with associated phonemes (set **A**). The green checkmarks indicate commutative paths in the olog where starting from the same instance of a set two distinct paths point to the same instance of another set. The way the olog is represented here correlates one-to-one to a computer implementation. **c**, A hierarchical olog yields identical information but a better overview of the underlying structure of the problem.
11

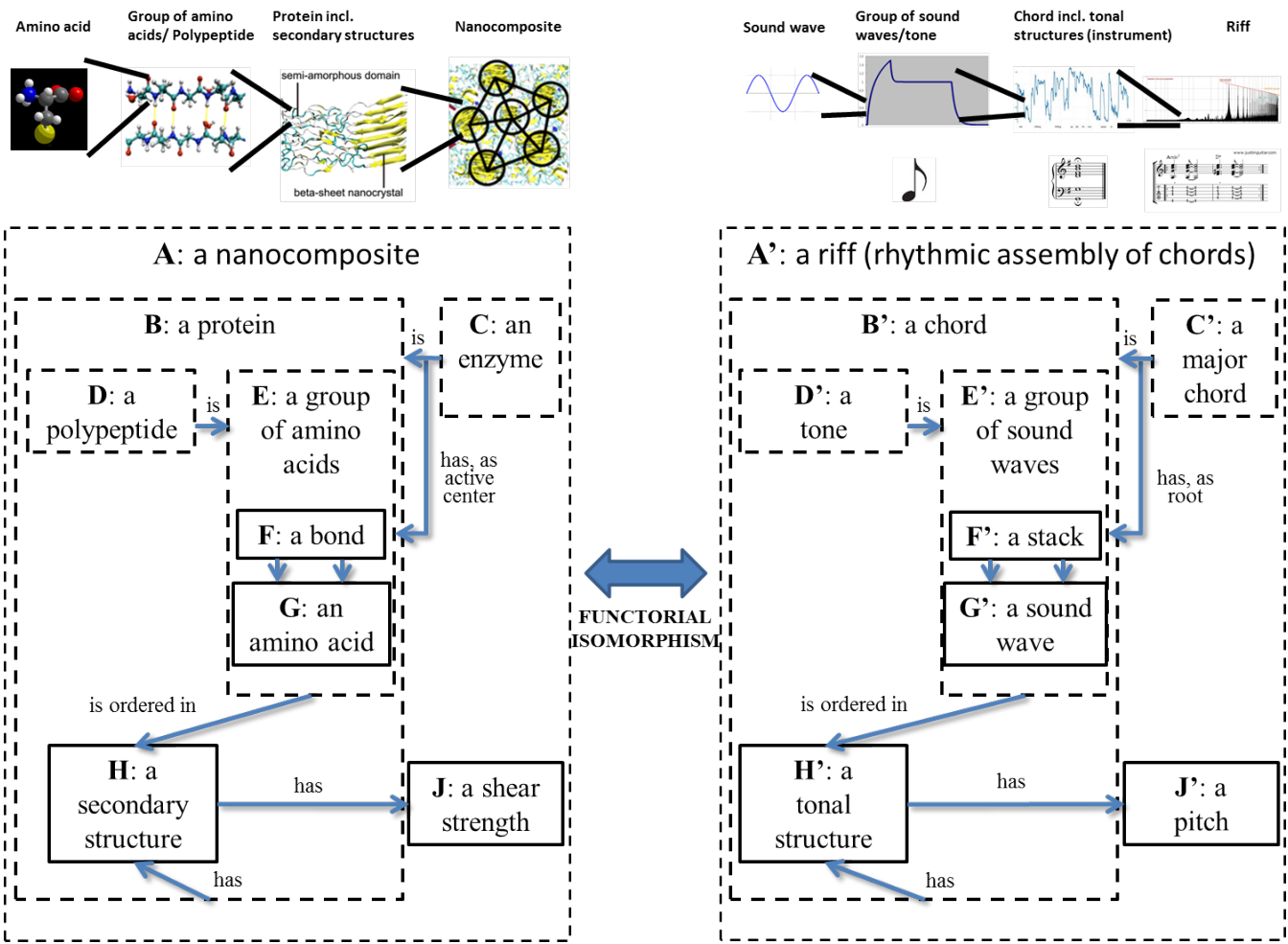

**Figure 2 | Functorial analogy of protein materials and musical structure by hierarchical ologs.** (top)The different hierarchical levels in spider silk and music are identified via experiment, theoretical analysis and computational studies. (left) Olog describing the hierarchical buildup of proteins out of their building blocks, amino acids. Each bond has as starting point and end point an amino acid. A polypeptide is a linear chain of two or more amino acids connected by bonds. Nanocomposites consist of proteins in certain secondary structures (*e.g.* α-helix, β-sheet or amorphous phase; in any case defined by a graph structure) of a specific size that assemble into higher level networks. The geometry of these of the secondary structure directly relates to macroscopic functional properties such as shear strength and extensibility. Each hierarchy-box represents a state or 'snapshot' of its inner constituents that are connected in a graph structure, as shown in Figure 1. This assures the unique allocation of lower level elements to superior structures. Apart from the hierarchical levels, a subset of proteins, the enzymes, are included to exemplify the way to include 'hierarchical interaction'. A function relates a certain property of a higher hierarchical structure, here active centers of enzymes, to an element of a lower level structure, a group of amino acids. (right) A *functorial isomorphism* relates objects and arrows from the protein network to a music network which shows an equivalent hierarchical buildup. The building blocks in music networks are basic sound waves (sine, triangular, sawtooth, *etc.*) that assemble via stacking. The variation of frequency and amplitude of the stacked waves leads to the formation of the functional unit 'a note' defined by its property 'a pitch'. Similar to the enzyme-protein relation certain subsets of chords, here the major chords, include group members with distinguished functional meaning.

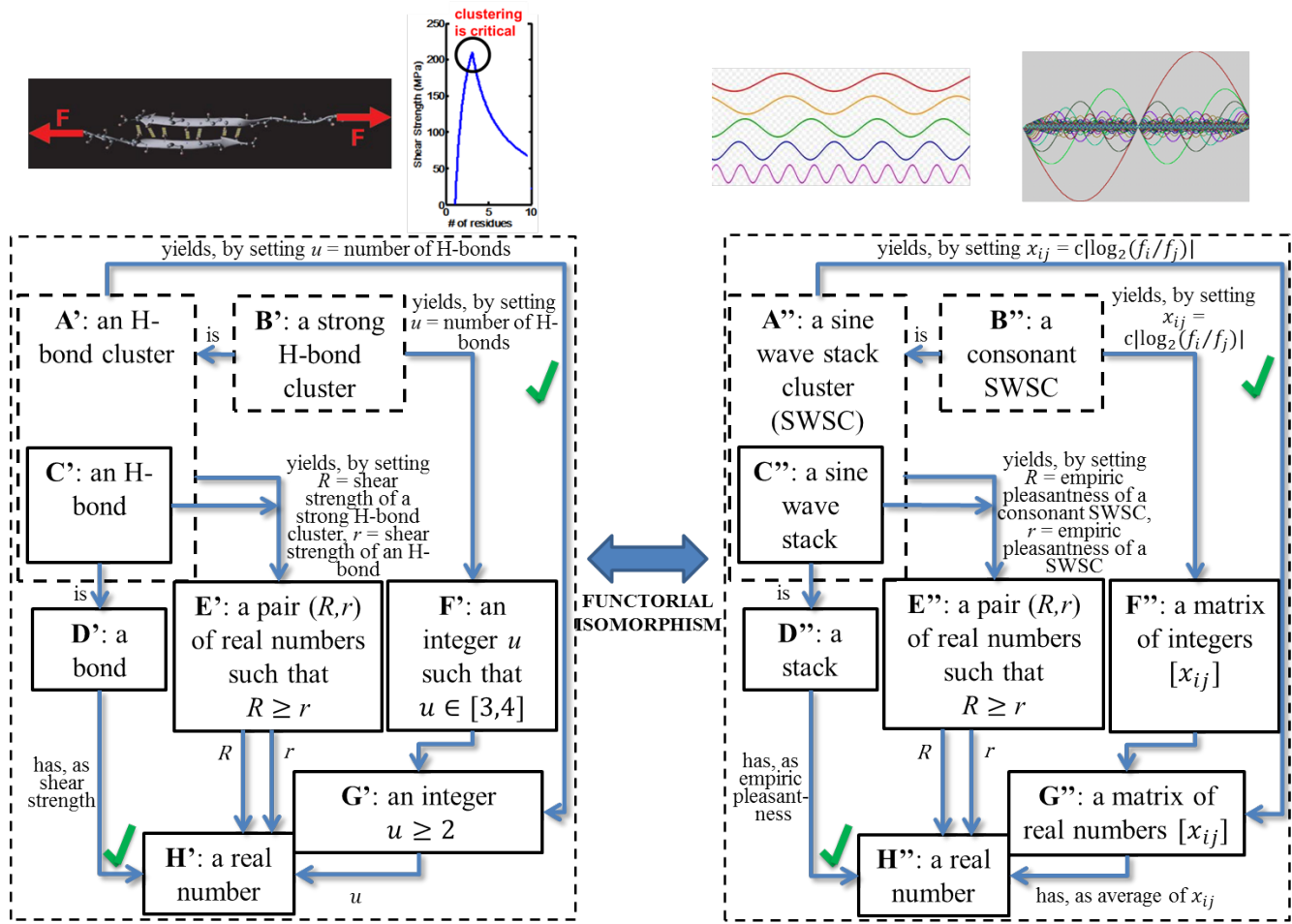

**Figure 3 | Functorial analogy of protein materials and musical function by hierarchical ologs.** (left) Olog describing the dependence of clustering to obtain functionality. For protein materials such as spider silk hydrogen bonds (H-bonds) cluster into groups of 3-4 residues. Compared to the strength of a single bond, the shear strength of such a bond cluster becomes significantly higher. In the olog this is typically modeled by relations of real numbers. Therefore, the property must be quantitatively ascertainable. (right) Analogous to this, chords in music form by stacking sound waves on an equal tempered scale, *i.e.* with frequencies that have integer ratios. Unlike shear strength the benefit of sound wave clustering – in so called consonant cluster - is not easily quantifiable but is subject to empirical observations, the measured pleasantness. The condition for consonance is given by the condition that all entrances of the frequency matrix are integers. Note, that the paths C→D→H and C→E→H as well as the paths B→A→G and B→F→G commute. Both ologs can be adjoined to the ologs shown in Figure 2 simply by connecting it to the set 'a bond'/'a stack'.